\documentclass[aps,prl,twocolumn,nofootinbib,superscriptaddress,floatfix]{revtex4-1}  
\usepackage{array,multirow,graphicx}  

\usepackage{amsmath}
\usepackage{amssymb}   
\usepackage{color}
\usepackage{hyperref}
\usepackage{ulem}
\usepackage{natbib}
\usepackage{enumitem}

\newcommand{\ltsima} {$\; \buildrel < \over \sim \;$}
\newcommand{\gtsima} {$\; \buildrel > \over \sim \;$}
\newcommand{\lta} {\lower.5ex\hbox{\ltsima}}
\newcommand{\gta} {\lower.5ex\hbox{\gtsima}}

\def\ln{\mathrm{ln}}

\newcommand{\vtheta}{{\boldsymbol{\theta}}}

\newcommand{\post}{\tilde p(\vtheta|\vx,M)}

\newcommand{\boldvec}[1]{{\mbox{\boldmath{$#1$}}}}
\providecommand{\Planck}{\textit{Planck}}
\providecommand{\Omb}{\Omega_{\mathrm{b}}}
\providecommand{\omb}{\omega_{\mathrm{b}}}
\providecommand{\Omc}{\Omega_{\mathrm{c}}}
\providecommand{\omc}{\omega_{\mathrm{c}}}
\providecommand{\Omk}{\Omega_K}

\providecommand{\Omtot}{\Omega_{\mathrm{tot}}}
\newcommand{\vx}{\boldvec{x}}
\newcommand\be{\begin{equation}}
\newcommand\ee{\end{equation}}
\newcommand{\rstar}{r_{\ast}}
\newcommand{\As}{A_{\rm s}}
\newcommand{\Mpc}{\text{Mpc}}
\newcommand{\ns}{n_{\rm s}}
\newcommand{\nrun}{d \ns / d\ln k}

\newcommand{\yhe}{Y_{\text{P}}}

\newcommand{\neff}{N_{\rm eff}}
\newcommand{\mnu}{\sum m_\nu}

\newcommand{\mnusterile}{m_{\nu,\, \mathrm{sterile}}^{\mathrm{eff}}}

\newcommand{\Alens}{A_{\rm L}}
\newcommand{\Aphiphi}{A_{\rm L}^{\phi\phi}}

\newcommand{\rpivot}{r_{0.05}}

\begin{document}

\title{No evidence for extensions to the standard cosmological model}

\author{Alan Heavens}\email{a.heavens@imperial.ac.uk}
\affiliation{Imperial Centre for Inference and Cosmology (ICIC), Imperial College, Blackett Laboratory, Prince Consort Road, London SW7 2AZ, U.K.}
\author{Yabebal Fantaye}
\affiliation{African Institute for Mathematical Sciences,
68 Melrose Road,
Muizenberg 7945,
South Africa}
\affiliation{Department of Mathematics, University of Stellenbosch, Stellenbosch 7602, South Africa}
\author{Elena Sellentin}
\affiliation{Imperial Centre for Inference and Cosmology (ICIC), Imperial College, Blackett Laboratory, Prince Consort Road, London SW7 2AZ, U.K.}
\affiliation{D\' epartement de Physique Th\' eorique,
Universit\' e de Gen\` eve,
Quai Ernest-Ansermet 24
CH-1211 Gen\` eve, Switzerland}
\author{Hans Eggers}
\affiliation{Department of Physics,
Stellenbosch University,
P/Bag X1,
7602 Matieland, South Africa }
\affiliation{National Institute for Theoretical Physics, Stellenbosch, South Africa}
\author{Zafiirah Hosenie}
\affiliation{Centre for Space Research, North-West University, Potchefstroom 2520, South Africa}
\affiliation{African Institute for Mathematical Sciences,
68 Melrose Road,
Muizenberg 7945,
South Africa}
\affiliation{South African Astronomical Observatory, Observatory Road, Observatory, Cape Town, 7935,South Africa}
\author{Steve Kroon}
\affiliation{CSIR-SU Centre for AI Research,
Computer Science Division,
Stellenbosch University,
P/Bag X1,
7602 Matieland, South Africa }
\author{Arrykrishna Mootoovaloo}
\affiliation{Department of Mathematics and Applied Mathematics, University of Cape Town,
Rondebosch, Cape Town, 7700, South Africa}
\affiliation{African Institute for Mathematical Sciences,
6–8 Melrose Road,
Muizenberg 7945,
South Africa}
\affiliation{South African Astronomical Observatory, Observatory Road, Observatory, Cape Town, 7935,South Africa}

\date{\today}

\begin{abstract}
We compute the Bayesian Evidence for models considered in the main analysis of \Planck\ cosmic microwave background data.  By utilising carefully-defined nearest-neighbour distances in parameter space, we reuse the Monte Carlo Markov Chains already produced for parameter inference to compute Bayes factors $B$ for many different model-dataset combinations.  Standard 6-parameter flat $\Lambda$CDM model is favoured over all other models considered, with curvature being mildly favoured only when CMB lensing is not included.  Many alternative models are strongly disfavoured by the data, including primordial correlated isocurvature models ($\ln B=-7.8$), non-zero scalar-to-tensor ratio ($\ln B=-4.3$), running of the spectral index ($\ln B = -4.7$), curvature ($\ln B=-3.6$), non-standard numbers of neutrinos ($\ln B=-3.1$), non-standard neutrino masses ($\ln B=-3.2$), non-standard lensing potential ($\ln B=-4.6$), evolving dark energy ($\ln B=-3.2$), sterile neutrinos ($\ln B=-6.9$), and extra sterile neutrinos with a non-zero scalar-to-tensor ratio ($\ln B=-10.8$).  Other models are less strongly disfavoured with respect to flat $\Lambda$CDM.   As with all analyses based on Bayesian Evidence, the final numbers depend on the widths of the parameter priors. We adopt the priors used in the \Planck\ analysis, while performing a prior sensitivity analysis.  Our quantitative conclusion is that extensions beyond the standard cosmological model are disfavoured by Planck data.  Only when newer Hubble constant measurements are included does $\Lambda$CDM become disfavoured, and only mildly, compared with a dynamical dark energy model ($\ln B\sim +2$).
\end{abstract}



\maketitle

\section{Introduction}

The standard cosmological model of flat $\Lambda$CDM is a remarkably simple and successful description of the Universe.  Based on cold dark matter (CDM) and a cosmological constant $\Lambda$, this flat model has 6 free parameters, which the \Planck\ satellite has measured with very high precision.  Extensions of the standard model have also been introduced, to relieve tensions with other datasets that have arisen with the standard model, or to probe for new physics, and in this respect model comparison is of more fundamental interest than parameter inference. Bayesian Evidence (or marginal likelihood) is the Bayesian tool to address such questions, and it can be challenging to compute as it requires integration of the likelihood over the multi-dimensional parameter space. In a companion paper \cite{Heavens2017} we show how Monte Carlo Markov Chains (MCMCs), produced for parameter inference, can be used to perform model comparison. In this letter, we report an analysis of all the main published  \Planck\ chains.  Bayesian Evidence has been computed for small numbers of models, e.g., by \cite{HB,Vennin} for curvaton models, \cite{Afshordi} for holographic models, \cite{MRV} for a comprehensive study of inflation models, \cite{BA} for inflationary features, \cite{FPV, LPV, Simpson}, who focussed on neutrino extensions. However this is the first comprehensive study of the models and datasets considered by \Planck.

\section{Bayesian Evidence}

The goal of parameter inference is to determine the posterior probability of model parameters $\vtheta$, given a data set $\vx$, any prior information and a model $M$. Using Bayes theorem, this is
\be
p(\vtheta|\vx, M) = \frac{p(\vx|\vtheta, M)\,\pi(\vtheta|M)}{p(\vx|M)}
\ee
where $\pi$ is the prior and $p(\vx|\vtheta, M)$ is the likelihood, which is regarded as a function of $\vtheta$. The Bayesian Evidence $p(\vx|M)$ is used for model comparison, and is the integral over the unnormalised posterior $\post \equiv p(\vx|\vtheta,M)\,\pi(\vtheta|M)$.  It trades the typically higher likelihood of the more complex model against the increased prior volume: 
\be
p(\vx|M) = \int d\vtheta\, p(\vx|\vtheta,M)\,\pi(\vtheta|M).
\ee
The posterior probability of competing models is then given by the product of the ratio of the model priors and the ratio of Evidences (the latter being known as the Bayes factor):
\be
\frac{p(M_1|\vx)}{p(M_2|\vx)} = \frac{\pi(M_1)}{\pi(M_2)}\,\frac{p(\vx| M_1)}{p(\vx|M_2)}.
\ee
The Evidence may be expensive to compute if the dimensionality of the parameter space is large.  Also we typically do not know $\tilde p$, but have only samples of it obtained by MCMC techniques.  Many such chains exist for the \Planck\ data, for various dataset-model combinations.   In this application the standard model is a special case of the extended models, so the maximum likelihood of the extended model will be at least as high as the standard model, so it is of limited use, whereas the Bayesian Evidence, which includes an element of Occam's razor, will quantify whether the increase of likelihood throughout the parameter space warrants support for the more complex model. 

Conventional wisdom is that MCMC chains are not good for computing the evidence, as it is claimed that they do not explore the tails of the distributions well.  i.e. marginalising over all parameters is thought not to be possible.  However, it is common to marginalise over all but one or two parameters, to obtain marginal posteriors for parameters individually or in pairs, and the tails do not seem to be a concern in those cases.  The real issue with Bayesian Evidence is the normalisation of the integral.  The MCMC samples from the unnormalised posterior, i.e. the chain is a sample from a number density $n(\vtheta)$ that is proportional to the unnormalised posterior, $\tilde p=a n $ but with an unknown constant of proportionality $a$.  If this constant can be determined, then the Evidence is readily computed, since, replacing $n$ by the sample density (a sum of Dirac delta functions),
\be
E = \int\,d\vtheta \,a\, n(\vtheta) = a \int\,d\vtheta\, \sum_{\alpha=1}^N \delta(\vtheta-\vtheta_\alpha) = a\,N
\ee
where $N$ is the length of the chain.  Alternatively, since $n=N p(\vtheta|\vx,M) = N \tilde p/E$, $E=N\tilde p/n \equiv aN$. 

We use the MCEvidence algorithm presented in \cite{Heavens2017}, where the nearest-neighbour distances $D$, in an $m$-dimensional MCMC chain are used in a Bayesian analysis to infer $a$.   The chain is pre-whitened such that the covariance matrix of the points is the identity, and a Euclidean distance measure then employed. This is equivalent to using the Mahalanobis distance, where the inverse covariance matrix defines the metric. The resulting posterior for the Bayesian Evidence $E$ (assuming a $1/E$ prior) is given by a sum over the MCMC points $\alpha$, each weighted by $w_\alpha$:
\be
\ln p(E| \{D_\alpha\}) =  {\rm const.} - (N+1)\ln E -\frac{W}{E}\sum_{\alpha=1}^N \frac{V_m(D_{\alpha}) \tilde p_\alpha}{w_\alpha},
\ee
where $V_m(D) = \pi^{m/2} D^m/\Gamma(1+m/2)$ is the volume of a $m-$ball of radius $D$ and $W\equiv \sum_\alpha w_\alpha$. We assume that the MCMC algorithm produces independent distances, but we test this later.  See \cite{Heavens2017} for more details.

The maximum a posteriori value of $E$ is
\be
E_{\rm MAP} = \frac{W}{N+1}\,{\sum_{\alpha=1}^N \frac{V_m(D_{\alpha})\tilde p_\alpha}{w_\alpha}},
\label{EMAP}
\ee
with a statistical variance in $\ln E$ of $1/(N+1)$.  We marginalise over nuisance parameters, and run a nearest-neighbour distance algorithm to determine $D_\alpha$, then compute the posterior for the Evidence.  Marginalising over nuisance parameters adds scatter to $\tilde p$, which increases the statistical error on $E$.  We have checked the effect of including some nuisance parameters and most $\ln B$ values change by $<0.1$.  Larger changes (up to about unity) occur only when $|\ln B|$ is itself very large.
\begin{table*}[tb] 
\begingroup 
\caption{Cosmological parameters (adapted from \cite{Planck13XVI}), their prior range in square brackets, the baseline values assumed, the nomenclature used in the model extensions in Table II, and a summary definition (see text for details).  The flat $\Lambda$CDM base model is parametrised by the parameters above the horizontal line. For completeness priors are given for all relevant extensions, even if the models are not discussed in this letter.}
\label{tab:params}
\footnotesize 

\begin{tabular}{lccll}
\hline
\newdimen\digitwidth 
\setbox0=\hbox{\rm 0}
\digitwidth=\wd0
\catcode`*=\active
\def*{\kern\digitwidth}
\newdimen\signwidth
\setbox0=\hbox{+}
\signwidth=\wd0
\catcode`!=\active
\def!{\kern\signwidth}
Parameter & Prior range & Baseline & Nomenclature & Definition\cr
\hline
$\omb \equiv \Omb h^2$& $[0.005, 0.1]$ & \dots&  &Baryon density today\cr
$\omc \equiv \Omc h^2$& $[0.001, 0.99]$& \dots& & Cold dark matter density today\cr
$100\theta_{\mathrm{MC}}$ & $[0.5, 10]$ & \dots& & $100\,{\times}$ approximation to $\rstar/D_{\rm A}$ (CosmoMC)\cr
$\tau                $&   $[0.01, 0.8]$ & \dots& & Thomson scattering optical depth due to reionization\cr
$\ln(10^{10}\As) $& $[2, 4]$ & \dots& & Log power of the primordial curvature perturbations ($k_0 = 0.05\,\Mpc^{-1}$)\cr
$\ns           $& $[0.8, 1.2]$ & \dots& &Scalar spectrum power-law index ($k_0 = 0.05\Mpc^{-1}$)\cr
\hline
$\Omk            $&  $[-0.3, 0.3]$ & 0& omegak &Curvature parameter today with $\Omtot= 1 - \Omk$\cr
$\mnu        $& $[0, 5]$ & $0.06$ & mnu &The sum of neutrino masses in eV\cr
$\mnusterile$&  $[0, 3]$ &0& meffsterile &Effective mass of sterile neutrino in eV\cr
$w_0                $& $[-3, 1]$ & $-1$& w&Dark energy equation of state, $w(a) = w_0 + (1-a) w_a$\cr
$w_a                 $& $[-3, 2]$ & 0&  w\_wa &As above (perturbations modelled using PPF)\cr
$\neff       $& $[0.05, 10]$ & 3.046& nnu & Effective number of neutrino-like relativistic degrees of freedom\cr
$\yhe                $& $[0.1, 0.5]$ & BBN& yhe & Fraction of baryonic mass in helium\cr
$\alpha_{-1}        $& [-1,1] & 0& alpha1 &Correlated isocurvature parameter\cr
$\Alens        $& $[0,  10]$& 1& Alens &Amplitude of the lensing power relative to the physical value\cr
$\Aphiphi           $& [0,10] & 1& Aphiphi & Amplitude of the lensing power from the 4-point function relative to the physical value\cr
$\nrun$&   $[-1, 1]$ & 0& nrun &Running of the spectral index\cr
$\rpivot          $& $[0, 3]$ & 0& r &Ratio of tensor primordial power to curvature power at $k_0 = 0.05\,\Mpc^{-1}$\cr
\hline
\end{tabular}
\endgroup
\end{table*}

\section{Data and Models}

The \Planck\ chains use a variety of datasets, which are detailed in \cite{Planck13XVI}.

The models are the `base' flat $\Lambda$CDM model, with parameters $\omega_{\rm b}=\Omega_{\rm b} h^2$, $\theta_{\rm MC}$, $\omega_{\rm c}=\Omega_{\rm c} h^2$, $\tau$, $\ln(10^{10}A_{\rm s})$ and $n_{\rm s}$, representing baryon density, CDM density, angle of the first peak, optical depth, amplitude of primordial fluctuations, and scalar spectral index.  $h$ is the Hubble parameter $H_0/100$km$\,$s$^{-1}$Mpc$^{-1}$.  The other models consist of the base model with 1-3 extensions.  Extensions and prior ranges are listed in Table 1.  It takes about 15 minutes on a laptop to analyse all the chains with MCEvidence.

\section{Results}

Table II shows Bayes factors with respect to the favoured model for selected datasets.  Bayes factors for other \Planck\ datasets are available online.  Internal analysis indicates a typical statistical error of $\sim 0.02$. The scatter in the mean of the MAP estimates from individual chains is typically $0.02-0.1$.    In the revision \cite{KR} of the Jeffreys scale, $|\ln B|>3$ is regarded as strong evidence (relative probability $>20$), and $|\ln B|>5$ as very strong (relative probability $>150$).   As is always the case, the Bayesian Evidence depends on the priors chosen.  Here we have used uniform box priors based on\footnote{In addition, cosmoMC excludes some physically impossible subregions.} the 2015 \Planck\ analysis (Antony Lewis, priv. comm.), and these are listed in Table I. For parameter inference these are not important if the data are informative, and some are set to wide uniform priors.  In high dimensions the Evidence is very prior dependent, but Bayes factors depend only on the width of the prior of the (usually one) additional parameter(s).   $\ln B$ values can be adjusted to a new prior range $\Delta\theta$ (assuming it is larger than the extent of the chain), by adding $\ln(\Delta\theta_{\rm Table\ I}/\Delta\theta)$ to the table values.  A narrower prior range increases the relative probability of the extended model.  Some models are not excluded with high probability, and one should be cautious of these, given the prior uncertainty.  However, we see that \Planck\ data provide very strong evidence against some models\footnote{Note that in the text we discuss Bayes factors compared with flat $\Lambda$CDM.  In Table II the numbers quoted are w.r.t. the highest Evidence model, which for a few datasets is not the base model.}: e.g., correlated isocurvature models are disfavoured with $\ln B=-7.8$ by \Planck\ alone, using polarisation; running of the spectral index is disfavoured ($-4.7$); and the evidence is against nonstandard neutrino masses and number ($-6.5$). The evidence against non-zero $r_{0.05}$ is strong compared with the base model ($-4.3$).

\begin{figure}
\includegraphics[width=8cm,angle=0]{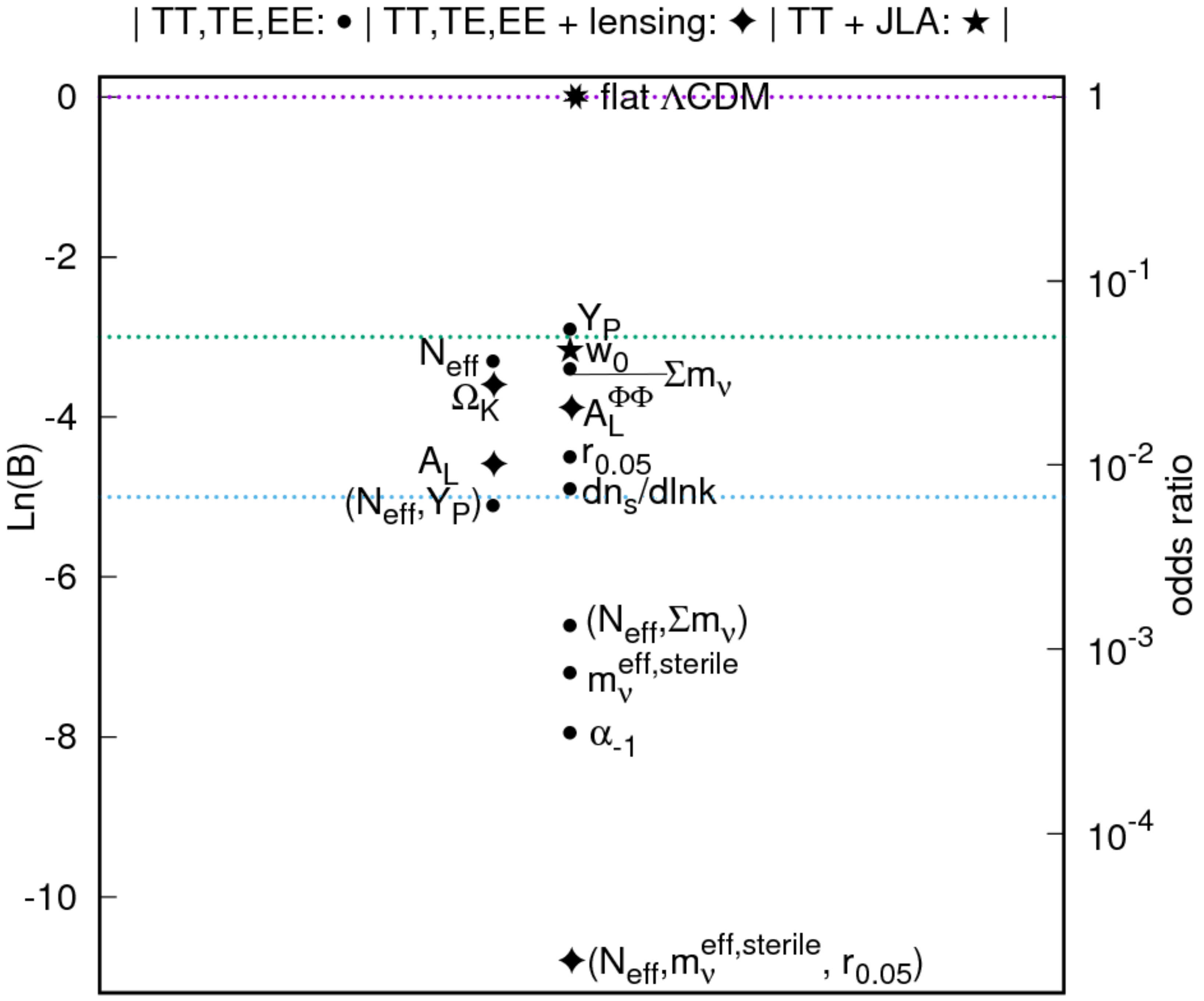}
\caption{Bayes factors $\ln B$ w.r.t. the highest evidence model (base:  flat $\Lambda$CDM).  The most constraining dataset is indicated by the symbol; see legend for details.
Horizontal lines mark the boundaries corresponding to strong ($\ln B<-3$) and very strong ($<-5$) evidence in the Kass \& Raftery \cite{KR} scale. }
\label{ResultsPlot}
\end{figure}

At a weaker level,  we find that a non-flat Universe is marginally preferred over the standard model by \Planck\ data until CMB lensing is included, after which there is strong evidence in favour of the flat standard model ($-3.6$).  There is no evidence for non-standard lensing ($-4.6$), or for varying the number of neutrino species ($-3.1$) or masses ($-3.2$).  A model including massive extra neutrinos, which was introduced to alleviate tension with direct Hubble constant measurements (e.g. \cite{Riess2016}) and low-amplitude cosmic shear (e.g.  \cite{HH,VPJ,Wyman,KiDS450}) has a very low probability from Planck data alone ($\ln B = -6.9$ w.r.t. base), and a model which also has non-zero tensor-to-scalar ratio $r$ has exceptionally strong evidence against ($-10.8$).   The results are shown in Fig. \ref{ResultsPlot}.

These results are understood when compared with the marginal distributions of parameters from the \Planck\ chains. For example, in the absence of CMB lensing the posterior for $\Omega_m$ and $\Omega_\Lambda$ follows the geometric degeneracy line (Fig. 26 of \cite{Planck15XIII}), with most of the probability lying away from the flat line in this plane.  When CMB lensing is included, the posterior is concentrated close to the intersection of the lines, and the Bayesian Evidence favours the flat model.  Correlated isocurvature modes are similarly constrained to be very close to zero amplitude (Fig. 24 of \cite{Planck15XIII}), when TE and EE polarisation is included.   Similar observations can be made for other parameters.  Note that naive interpretations of credible intervals may not be supported by the Evidence. One example of this is the lensing power amplitude, for which $A_L=1$ is in the tails of the distribution for some datasets (see Fig. 12 of \cite{Planck15XIII}), but for which the evidence favours the standard model. See Fig. 3 of \cite{Trotta} for illustration of this general point.


\begin{table*}[htbp]
  \centering
  \caption{Bayes factors with respect to the model favoured by each dataset (flat $\Lambda$CDM except for Planck-only without lensing).   For full description of datasets, see the \Planck\ descriptions, where names should be preceded by {\bf base$\_$plikHM$\_$}; in short BAO=baryon acoustic oscillations, lensing=CMB lensing, JLA=supernovae, H070p6=Hubble constant prior centred on 70.6, zre6p5=recombination at $z=6.5$. For model nomenclature, see Table I. Bayes factors in bold are referred to in the text, adjusted where necessary if the $\Lambda$CDM Evidence is not the highest in the column (designated by italics). This is a subset of the model/dataset combinations considered in this analysis.  Full results are available from http://astro.ic.ac.uk/aheavens/home.}
        \begin{tabular}{|r|l|r|r|r|r|r|r|r|r|r|r|r|r|r|r|r|r|r|r|r|}
        \hline
    & $\ln\,B$ & \rotatebox[origin=l]{90}{\textbf{TT\_lowTEB}} & \rotatebox[origin=l]{90}{\textbf{TT\_lowTEB\_post\_BAO}} & \rotatebox[origin=l]{90}{\textbf{TT\_lowTEB\_post\_lensing}} & \rotatebox[origin=l]{90}{\textbf{TT\_lowTEB\_post\_H070p6}} & \rotatebox[origin=l]{90}{\textbf{TT\_lowTEB\_post\_JLA}} & \rotatebox[origin=l]{90}{\textbf{TT\_lowTEB\_post\_zre6p5}} & 
\rotatebox[origin=l]{90}{\textbf{TT\_lowTEB\_BAO}} & \rotatebox[origin=l]{90}{\textbf{TT\_lowTEB\_BAO\_post\_lensing}} & \rotatebox[origin=l]{90}{\textbf{TT\_lowTEB\_lensing}} & \rotatebox[origin=l]{90}{\textbf{TT\_lowTEB\_lensing\_post\_BAO}} & \rotatebox[origin=l]{90}{\textbf{TT\_tau07}} & \rotatebox[origin=l]{90}{\textbf{TT\_lowTEB\_lensing\_BAO}} & \rotatebox[origin=l]{90}{\textbf{TTTEEE\_lowTEB}} & \rotatebox[origin=l]{90}{\textbf{TTTEEE\_lowTEB\_post\_BAO}} & \rotatebox[origin=l]{90}{\textbf{TTTEEE\_lowTEB\_post\_lensing}} & \rotatebox[origin=l]{90}{\textbf{TTTEEE\_lowTEB\_post\_H070p6}} & \rotatebox[origin=l]{90}{\textbf{TTTEEE\_lowTEB\_post\_JLA}} & \rotatebox[origin=l]{90}{\textbf{TTTEEE\_lowTEB\_post\_zre6p5}} & \rotatebox[origin=l]{90}{\textbf{TTTEEE\_lowTEB\_lensing}} 
\\
    \hline
    1 & \textbf{base} & {\it -0.7}  & 0.0   &       & 0.0   & 0.0   & 0.0   &             &       & 0.0   & 0.0   & 0.0   &       & {\it -0.2}  & 0.0   &       & 0.0   & 0.0   & 0.0   & 0.0    \\
    2 & \textbf{omegak} & 0.0   &       &       &       &       &       &        -1.7  & -1.5  & -3.6  &       &       &       & 0.0   &       &       &       &        & &  {\bf -3.6} \\
    3 &\textbf{Alens} & -1.7  & -0.9  &       & -0.3  & -0.5  & -0.8         &       &       & -4.2  & -4.1  & -2.5  &       & -2.1  & -1.5  &       & -1.5  & -1.5  & -1.8  & {\bf -4.6}   \\
  4 &  \textbf{nnu} & -3.4  &       & -1.7  & -2.5  & -2.5  & -2.5         & 0.0   & 0.0   &       &       & -2.2  &       & {\bf -3.3}  &       & -1.7  & -3.0  & -2.9  & -2.9  &         \\
  5 & \textbf{mnu} & -3.7  &       &       & -3.2  & -3.2  & -2.8         & -0.5  &       & -2.6  &       &       & 0.0   & {\bf -3.4}  &       &       & -3.2  & -3.2  & -3.0  & -2.9    \\
  6 & \textbf{nrun} & -5.1  & -4.1  & -3.8  & -4.0  & -4.0  & -4.1   &       &       &       &       & -4.4  &       & {\bf -4.9}  & -4.5  & -3.8  & -4.5  & -4.5  & -4.5  &         \\
  7 &   \textbf{r} & -4.9  & -4.0  &       & -4.0  & -4.0  & -4.0       &       &       & -4.0  & -3.9  &       &       & {\bf -4.5}  & -4.1  &       & -4.1  & -4.1  & -4.1  & -4.1   \\
  8 &   \textbf{w} & -1.2  &       & 0.0   & -1.7  & {\bf -3.2}  & -0.2         & -0.3  & -0.2  &       &       &       &       & -0.7  &       & 0.0   & -1.9  &       & -0.3  &       \\
 9 &   \textbf{alpha1} & -6.4  & -5.6  & -5.1  & -5.5  & -5.5  & -5.4    &       &       &       &       &       &       & {\bf -8.0}  & -7.6  & -6.2  & -7.5  & -7.6  & -7.6  &        \\
   10& \textbf{Aphiphi} &       &       &       &       &       &       &              &       & -3.9  &       &       &       &       &       &       &       &       &       & -3.8    \\
11&  \textbf{yhe} & -2.9  & -2.0  & -1.1  & -1.8  & -1.9  & -1.9     &       &       &       &       &       &       & -2.9  & -2.5  & -1.5  & -2.5  & -2.5  & -2.5  &         \\
 12&   \textbf{mnu\_Alens} &       &       &       &       &            &       &       &       &       &       &       & -3.6  &       &       &       &       &       &       &         \\
  13&  \textbf{mnu\_omegak} &       &       &       &       &              &       &       &       &       &       &       & -4.9  &       &       &       &       &       &       &         \\
  14&  \textbf{mnu\_w} &       &       &       &       &       &              &       &       &       &       &       & -3.1  &       &       &       &       &       &       &         \\
   15& \textbf{nnu\_mnu} & -6.6  & -6.1  &       & -5.7         &       &       &       &       & -5.4  & -6.0  &       &       & -6.6  & -6.5  &       & -6.3  &       &       & -6.0   \\
  16&  \textbf{nnu\_yhe} & -5.2  & -4.4  & -3.4  & -4.2        &       &       &       &       &       &       &       &       & -5.1  & -4.6  & -3.3  & -4.8  &       &       &         \\
   17& \textbf{w\_wa} &       &       &       &       &              &       & -0.1  & -0.3  &       &       &       &       &       &       &       &       &       &       &         \\
   18& \textbf{nnu\_meffsterile} & -6.5  &       &       &              &       &       & -3.2  &       & -5.7  &       &       & -2.4  & {\bf -7.1}  &       &       &       &       &       & -6.6    \\
 19&   \textbf{nnu\_meffsterile\_r} &       &       &              &       &       &       &       &       & -9.9  &       &       &       &       &       &       &       &       &       & {\bf -10.8}   \\
    \hline
    \end{tabular}%
  \label{tab:addlabel}%
\end{table*}%

\subsection{Robustness tests}

MCEvidence assumes that the points in the chain are independent, and this is not strictly the case.  To test the effect of this we have computed the correlations to find that they are generally small, but we have also aggressively thinned the chains.  Thinning by a factor 10 makes little difference to the results, with the vast majority of $\ln B$ values changing by $< 0.2$, with very few over 0.5.  A few heavily disfavoured models change by more, up to 0.7, so the conclusions are robust.   We also note that if weights are set to unity in \Planck\ chains the same conclusions are reached.   

Since the Bayes factors depend on the width of the prior for the additional parameter(s), we can ask by how much they have to be changed for the models to be preferred over the standard model.  The most competitive models are disfavoured with $\ln B \simeq -3$, which requires the prior range to be reduced by a factor 20 for them to be more probable than flat $\Lambda$CDM.  For example, $w$ would need to be restricted to a prior range $<0.2$,  $Y_{\rm P}$ to 0.02, and $r_{0.05}$ to 0.03, within the current credible region.  The neutrino mass conclusion is least secure, as a prior range less than 0.2eV would favour non-standard masses.

\section{Discussion}

The main aim of this paper is to compute Bayesian Evidence values for the many models and datasets produced in the primary \Planck\ analysis, where we find that the 6-parameter flat $\Lambda$CDM model is preferred, with no evidence in favour of extensions.    As is usual with Evidence calculations, the results sometimes favour simpler models even when naive inspection of credible intervals suggests otherwise.   We agree with the conclusions of \cite{FPV,LPV}
but not \cite{BM} in disfavouring adding extra massive neutrino components to the base model, but our conclusions are far more wide-ranging.  We also complement the analysis of \cite{HJV} that shows a model-independent lack of evidence for deviations from standard physical parameters.   The inclusion of recent Hubble constant measurements \cite{Riess2016,Feeney}
(the latter with/without outliers) favours wCDM over $\Lambda$CDM, but only with modest odds ($\ln B=2.2,\ 2.0,\ 1.5$) respectively, from chains that allow all parameters to vary. We do not include strong lensing constraints on $H_0$ (e.g. \cite{Bonvin}) as the constraints are model-dependent so are not straightforward to add.

MCEvidence is written in Python and is freely available on Github at https://github.com/yabebalFantaye/MCEvidence.  The full table of Evidence results is linked from http://astro.ic.ac.uk/aheavens/home.

\paragraph{Acknowledgments.---}
We thank Andrew Jaffe, Bruce Bassett and Raul Jimenez for helpful discussions, and Antony Lewis for 2015 \Planck\ priors. YF is supported by the Robert Bosch Stiftung. ES is funded by a DAAD research fellowship of the German Academic Exchange Service. We thank referees for helpful comments.


\end{document}